\documentclass{ws-p8-50x6-00}
\usepackage[dvips]{color}       
\usepackage{epsfig}

\def\pmb#1{\setbox0=\hbox{#1}
\kern-.025em\copy0\kern-\wd0
\kern.05em\copy0\kern-\wd0
\kern-.025em\raise.0433em\box0}
\def\mbi#1{{\pmb{\mbox{\scriptsize ${#1}$}}}}

\def\bm#1{{\pmb{\mbox{${#1}$}}}}

\begin{document}

\def\beq{\begin{equation}}
\def\eeq{\end{equation}}
\def\beqa{\begin{eqnarray}}
\def\eeqa{\end{eqnarray}}
\def\d{{\rm d}}
\def\ttimes{{\scriptstyle \times}}
\def\half{{\textstyle {1\over2}}}

\title{Roper Electroproduction Amplitudes in a Chiral Confinement Model }

\author{M. Fiolhais, P. Alberto and  J. Marques}

\address{Departamento de F\'\i sica and Centro de F\'\i sica Computacional, 
Universidade de Coimbra, P-3004-516 Coimbra, Portugal\\ 
E-mail: tmanuel@teor.fis.uc.pt}

\author{B. Golli}

\address{Faculty of Education, University of Ljubljana  and
	         J. Stefan Institute, Ljubljana, Slovenia}

\maketitle

\abstracts{
A description of the Roper using the chiral chromodielectric model 
is presented and the transverse $A_{1/2}$ and the scalar $S_{1/2}$ 
helicity amplitudes for the electromagnetic Nucleon--Roper transition 
are obtained for small and moderate $Q^2$. The sign of the amplitudes 
is correct but the model predictions underestimate the data at the 
photon point. 
Our results do not indicate a change of sign
in any amplitudes up to $Q^2\sim1$~GeV$^2$. 
The contribution of the scalar meson excitations to the Roper 
electroproduction is taken into account but it turns out to be small 
in comparison with the quark contribution. 
However, it is argued that mesonic excitations may play a more 
prominent role in higher excited states.}

\section{Introduction}

Several properties of the nucleon and its excited states can be 
successfully explained in the framework of the constituent quark 
model (CQM), either in its non-relativistic or relativistic version.
There are, however, processes where the description in terms
of only valence quarks is not adequate
suggesting that other degrees of freedom may be important 
in the description of baryons, in particular the chiral mesons.
Typical examples -- apart of decay processes -- are electromagnetic 
and weak production amplitudes of the nucleon resonances. 
Already the production amplitudes for the lowest excited state,
the $\Delta$, indicate the important role of the pion cloud in the
baryons. 
The other well known example is
the Roper resonance, N(1440), which has been a challenge 
to any effective model of QCD at low or intermediate energies.  
Due to the relatively low excitation energy, a simple 
picture in which one quark populates the 2s level does not work. 
It has been suggested that the inclusion of  explicit 
excitations of gluons and/or glueballs, or explicit excitations 
of chiral mesons may be necessary to explain its properties. 

The other problem related to the CQM is the difficulty to introduce
consistently the electromagnetic and the axial currents as well
as the interaction with pions which is necessary to describe the
leading decay modes of resonances.
Such problems do not exist in relativistic 
quark models based on effective Lagrangians which incorporate 
properly the chiral symmetry.
Unfortunately, several chiral models for baryons, 
such as the linear sigma model or various versions of the 
Nambu--Jona-Las\'\i nio model,  
though able to describe properly the $\Delta$ resonance,
are simply not suited to 
describe higher excited states since they do not confine: 
for the nucleon, the three valence quarks in the lowest s
state are just bound and,  for typical parameter sets, 
the first radial quark excitation already lies in the continuum. 
In order to resolve this problem, other degrees
of freedom have to be introduced in the model to provide binding
also at higher excitation energies.
The chiral version of the chromodielectric model (CDM) seems to be
particularly suitable to describe radial excitations of the
nucleon since it contains the chiral mesons as well as a
mechanism for confining.
The CDM  has been used as a model for the nucleon~\cite{BirseP} 
in different approximations. 
Using the hedgehog coherent state approach supplemented 
by an angular momentum and isospin projection, 
several nucleon properties 
and of the nucleon-delta electromagnetic excitation have been 
obtained~\cite{BirseP,drago2,delta1}.

In the present work we concentrate on the description of the Roper
resonance. Its structure and the electroproduction amplitudes have
been considered in several versions of the CQM~\cite{Capstick,Cardarelli,Cano}.
The nature of the Roper resonance has also  been considered in a non-chiral 
version of 
the CDM using the RPA techniques to describe coupled vibrations 
of valence quarks and the background chromodielectric 
field~\cite{broRPA}.
The energy of the lowest excitation turned out to be
40~\% lower than the pure 1s--2s excitations.
A similar result was obtained by Guichon~\cite{guiMIT}, 
using the MIT bag model and considering the Roper as a 
collective vibration of valence quarks and the bag.

Our description of baryons in the framework of the CDM model
provides relatively simple model states 
which are straightforwardly used to compute the  transverse 
and scalar helicity amplitudes for the nucleon--Roper transition, 
in dependence of the photon virtuality~\cite{letter}. 
The electromagnetic probe (virtual photon) couples to charged 
particles, pions and quarks.  
However, in the CDM, baryons have got a weak pion cloud and 
therefore the main contribution to the electromagnetic nucleon--Roper  
amplitudes comes from the quarks.     

In Section~\ref{amplitudes} we introduce the
electromagnetic transition amplitudes.
In Section~\ref{model} we briefly describe the model and construct 
model states representing baryons, using the angular momentum 
projection technique from coherent states. 
In Section~\ref{resul}  we present the CDM predictions for the 
helicity amplitudes for typical model parameters. 
Finally, in Section~\ref{mesons} we discuss the contribution of 
scalar meson vibrations.

\section{Electroproduction amplitudes in chiral quark models 
\label{amplitudes}}

In chiral quark models the coupling of quarks to chiral fields
is written in the form:
\beq
    {\cal L}_{q-{\mbox{\scriptsize meson}}} = g\, \bar{q} \,
    (\hat{\sigma}+\mbox{i}\vec{\tau}\cdot\hat{\vec{\pi}}\gamma_5)\,q\;.
   \label{Lqm}
\eeq
Here $g$ is the coupling parameter related to the mass of the
constituent quark $M_q = gf_\pi$.
In the CDM the parameter $g$ is substituted by the 
{\em chromodielectric\/} field which takes care of the
quark confinement as explained in the next section.
In the linear $\sigma$-model, in the CDM, as well as in different
versions of the Cloudy Bag Model, 
the chiral meson fields, i.e. the isovector triplet of pion
fields, $\vec{\pi}$, and the isoscalar $\sigma$ field 
(not present in non-linear versions),  
are introduced as effective fields with their own dynamics
described by the meson part of the Lagrangian:
\beq
  {\cal L}_{\mbox{\scriptsize meson}} =
    \half\partial_\mu\hat{\sigma}\,\partial^\mu\hat{\sigma}
  + \half\partial_\mu\hat{\vec{\pi}}\cdot\partial^\mu\hat{\vec{\pi}}
  - {\cal U}(\hat{\sigma}, \hat{\vec{\pi}}) 
\label{Lm}
\eeq
where ${\cal U}$ is the Mexican-hat potential describing the
meson self-interaction.
In different versions of the Nambu--Jona-Las\'\i nio model~\cite{NJL}
the chiral fields are explicitly constructed in terms of quark-antiquark 
excitations of the vacuum in the presence of the valence quarks.

From (\ref{Lm}) and from the part of the Lagrangian corresponding to free
quarks, 
\beq
  {\cal L}_q = \mbox{i}\bar{ q  }\gamma^\mu \partial_\mu q\, ,
\label{Lq}
\eeq
the electromagnetic current is derived as the conserved Noether current:
\beq
 \widehat{J}_{e.m.}^\mu ({\bm r})  =
 \bar{ q  }\, \gamma^\mu \left( {\textstyle{1\over6}} +
\textstyle{1\over2}\tau_3\right) q
  + (\hat{\vec{\pi}}\times \partial^\mu \hat{\vec{\pi}})_3\, .
\label{charge}
\eeq
Note that the operator contains both the standard quark part
as well as the pion part.
We stress that in all these models the electromagnetic 
current operator is derived directly from the Lagrangian, hence 
no additional assumptions have to be introduced in 
the calculation of the electromagnetic amplitudes.

We can now readily write down the amplitudes for the electroexcitation
of nucleon excited states in terms of the EM current (\ref{charge}).
Let us denote by $|\tilde{\rm N}_{M, M_T }\rangle$ and 
$|\tilde{\rm R}_{J, T; M, M_T }\rangle$ the model states representing 
the nucleon and the resonant state, respectively (the indexes $M$ and $M_T$ 
stand for the angular momentum and isospin third components).
The resonant transverse  and scalar helicity amplitudes,
$A_\lambda$ and $S_{1/2}$ respectively, defined in the rest 
frame of the resonance, are
\begin{equation}
A_\lambda=- \zeta\,\sqrt{2 \pi \alpha \over k_W}  \int \d^3 {\bm r} \, \
  \langle\tilde{\rm R}_{J,T;\lambda, M_T}|{\bm J}_{\rm em}({\bm r})\cdot 
   {\bm \epsilon}_{+1} \, {\rm e} ^{{\rm i}{\mbi k} 
     \cdot {\mbi r} } |\tilde{\rm N}_{\lambda-1, M_T }\rangle
\label{helicea}
\end{equation}
\begin{equation}
S_{1/2}=\zeta\,\sqrt{2 \pi \alpha \over k_W}  \int \d {\bm r}  \, 
    \langle\tilde{\rm R}_{J,T;+{1\over 2}, M_T}|  J^0_{\rm em} ({\bm r}) \,\, 
     {\rm e} ^{{\rm i}{\mbi k} 
     \cdot {\mbi r} } |\tilde{\rm N}_{+{1\over 2}, M_T}\rangle\, ,
\label{helices}
\end{equation}
where $\alpha={e^2\over 4\pi}={1\over 137}$ is 
the fine-structure constant, 
the unit vector ${\bm \epsilon}_{+1}$ is the polarization vector of 
the electromagnetic field,  $k_W=(M_{\rm R}^2 - M_{\rm N}^2) / 2 M_{\rm R}$ 
is the photon energy at the photon point (introduced rather than 
$\omega$ which vanishes at $Q^2=M_R^2-M_N^2$) and $\zeta$ is the 
sign of the $N\pi$ decay amplitude. 
This sign has to be explicitly calculated within the model;
from (\ref{Lqm}) in our case.
In the case of the $\Delta$ resonance ($T=J={3\over2}$), 
$\lambda$ takes two values $\lambda={3\over2}$ and ${1\over2}$, 
while for the Roper state ($T=J={1\over2}$),
only one transverse amplitude exists ($\lambda={1\over2}$).

The photon four momentum is $q^\mu(\omega, {\bm k})$ and we define 
$Q^2=-q_\mu\, q^\mu$. In the chosen reference frame the following 
kinematical relations hold: 
\beq
\omega= {M_R^2-M_N^2-Q^2\over2M_R}\, ;
\qquad
{\bm k}^2 \equiv k^2=
\biggl[ {M_R^2+M_N^2+Q^2\over 2M_R} \biggr]^2-M_N^2\, .
 \label{ener}
\ee

The electroexcitation amplitudes for the $\Delta$ resonance have
been analyzed in the framework of chiral quark 
models~\cite{delta1,njle2m1,tiator}.
They are dominated by the M1 transition but contain also rather 
sizable quadrupole contributions E2 and C2.
The CQM model predicts here too low values for the M1 piece
and almost negligible values for the quadrupole amplitudes.
In chiral quark models there is a considerable contribution
from the pions (i.e. from the second term in (\ref{charge})): up to 50~\% in the M1 amplitude,
and they
dominate the E2 and C2 pieces.
The absolute values of the amplitudes and their behavior as a
function of the photon virtuality $Q^2$ is well reproduced
in the linear $\sigma$-model.
Though the ratios E2/M1 and C2/M1 are also well reproduced in the CDM,
this model gives systematically too low values for the amplitudes, which
could be attributed to its rather weak pion cloud.
As we shall see in Section~\ref{resul}, this might also explain the 
small values of the Roper production amplitudes at low $Q^2$.

In the next section we construct the states $|\tilde{\rm N}\rangle$ 
and $|\tilde{\rm R}\rangle$ for the Roper in the framework of the CDM and, 
in Section~\ref{resul}, we present the model predictions for the amplitudes.

\section{Baryons in the CDM \label{model}}

The Lagrangian of the CDM contains, apart of the chiral meson
fields $\sigma$ and $\pi$, the cromodielectric field $\chi$ such that
the quark meson-interaction (see (\ref{Lqm})) is modified as:
\beq
    {\cal L}_{q-{\mbox{\scriptsize meson}}} = {g\over\chi}\, \bar{ q  } \,
    (\hat{\sigma}+\mbox{i}\vec{\tau}\cdot\hat{\vec{\pi}}\gamma_5)\, q\;.
   \label{lagrangian2}
\eeq
The idea behind the introduction of the $\chi$ field is that
it acquires a nonzero expectation value inside the baryon but
goes to 0 for larger distances from the center of the baryon,
thus pushing the effective constituent quark mass to infinity
outside the baryon.
In addition, the Lagrangian  
contains kinetic and potential pieces for the
$\chi$-field: \beq
  {\cal L}_\chi =  \half\partial_\mu\hat{\chi}\,\partial^\mu\hat{\chi}
   - {1\over2}M_\chi^2\,\hat{\chi}^2  \, ,
  \label{lagrangian3}
\eeq
where $M_\chi$ is the $\chi$ mass.
In this work we consider only a simple quadratic potential;
other versions of the CDM assume more complicated forms, namely quartic 
potentials.

The free parameters of the model have been chosen by requiring
that the calculated static properties of the nucleon agree
best with the experimental values~\cite{drago2}.
In the version of the CDM with a quadratic potential, the results are
predominantly sensitive
to the quantity $G=\sqrt{gM_\chi}$; we take $G=0.2$~GeV
(and $g=0.03$~GeV).
The model contains  other parameters: the pion decay constant, 
$f_\pi=0.093$~GeV,
the pion mass, $m_\pi=0.14$~GeV, and the sigma
mass, which we take in the range 0.7~$\le m_\sigma\le1.2$~GeV.

The nucleon is constructed by placing three valence quarks
in the lowest s-state, 
i.e., the quark source can be written as (1s)$^3$. 
For the Roper the quark source is (1s)$^2$(2s)$^1$, i.e.~one of 
the three quarks now occupies the first (radially) excited state. 
The quarks are surrounded by a cloud of pions, sigma mesons
and chi field, described by radial profiles
$\phi(r)$, $\sigma(r)$ and $\chi(r)$ respectively.
The  hedgehog ansatz is assumed 
for the quarks and pions.
The quark profiles (described in terms of the upper, $u$, and
the lower component, $v$) and boson profiles are determined 
self-consistently.

Because of the hedgehog structure, the solution is neither an 
angular momentum eigenstate nor an isospin eigenstate, 
and therefore it cannot be related directly with a physical baryon. 
However, the physical states can be obtained from the hedgehog
by first interpreting the solution as a coherent state
of three types of bosons and then performing the 
Peierls-Yoccoz projection~\cite{BirseP,GR85}:
\begin{equation}
|{\rm N}_{{1\over 2},M_T}\rangle 
   = \mathcal{N}\,
     P^{1\over 2}_{{1\over 2},-M_T} |Hh\rangle \, , 
\quad 
    |{\rm R}'_{{1\over 2},M_T}\rangle 
     = \mathcal{N}'\,
     P^{1\over 2}_{{1\over 2},-M_T} |Hh^*\rangle \, ,   
\label{usa}
\end{equation}
where $P$ is the projector and we introduced the symbol * to 
denote the Roper intrinsic state.
Because of their trivial tensor nature, the $\chi$ and 
the $\sigma$-fields are not affected by projection.
This approach can be considerably improved by determining
the radial profiles $\phi(r)$, $\sigma(r)$ and $\chi(r)$, 
as well as  the quark profiles, 
using the variation after projection method~\cite{BirseP}, 
separately for the nucleon and for the Roper.

\begin{figure}[bht]
\centerline{\epsfig{clip=on,file=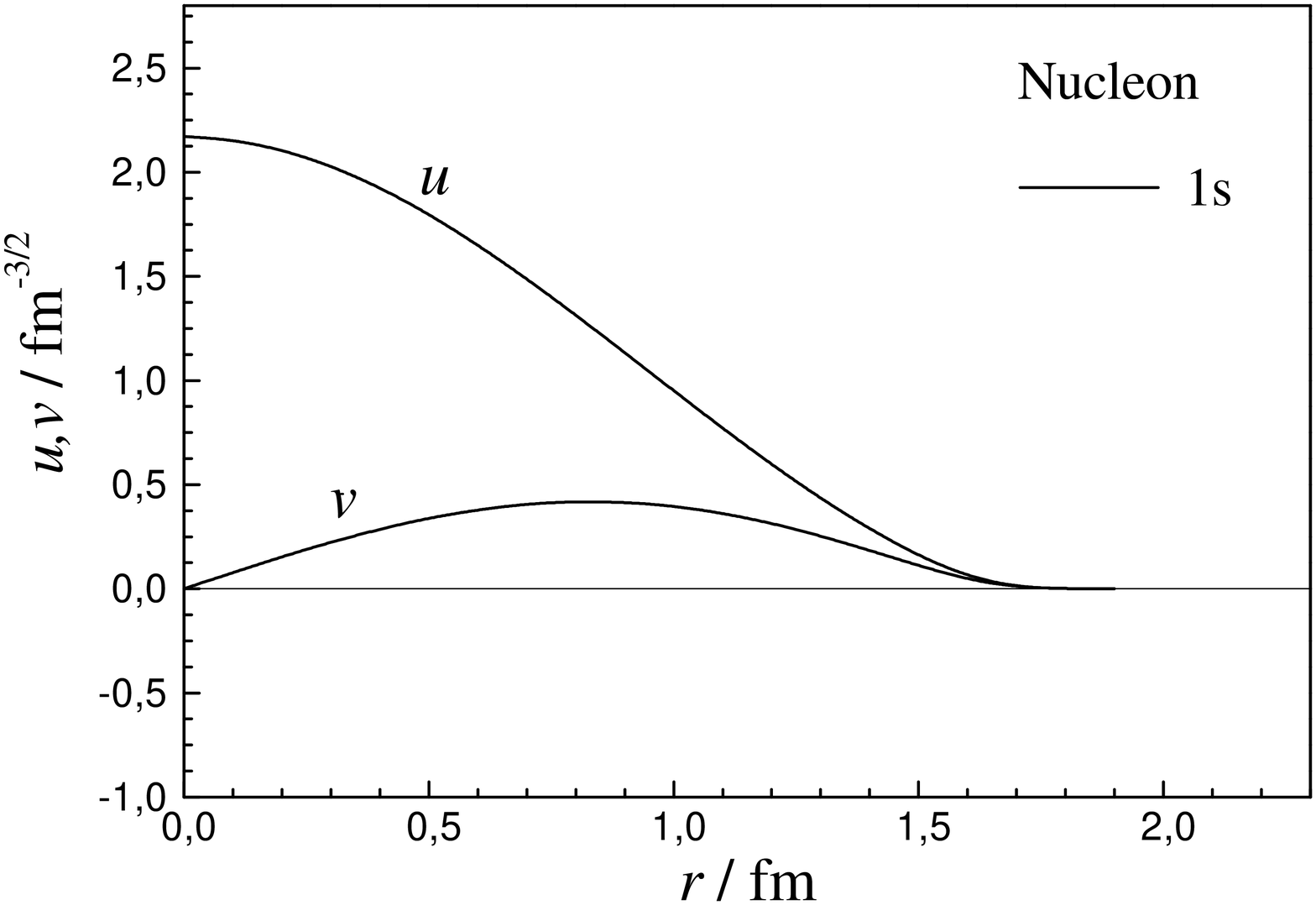,width=6cm}\epsfig{clip=on,
file=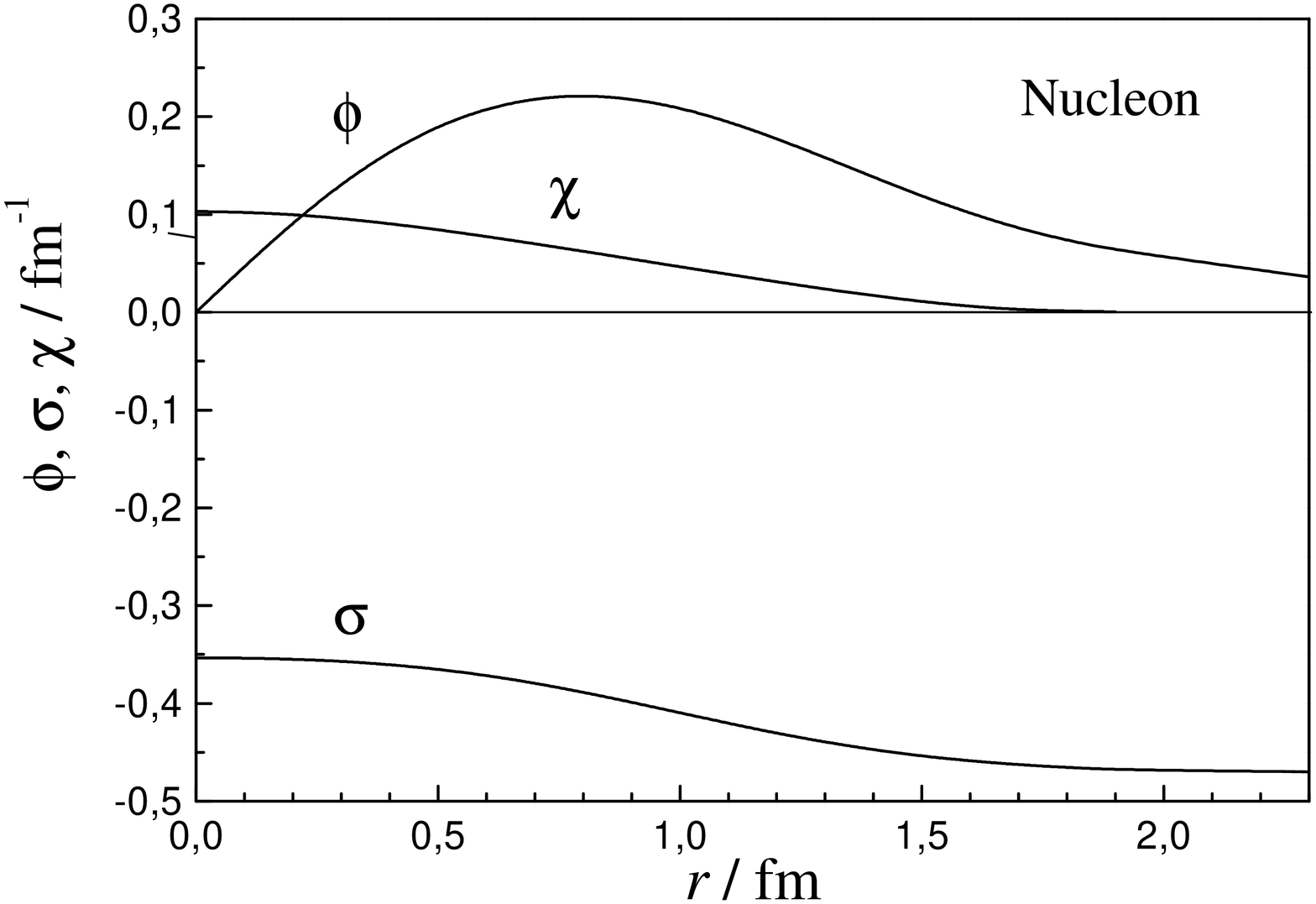,width=6cm}}
\centerline{\epsfig{clip=on,file=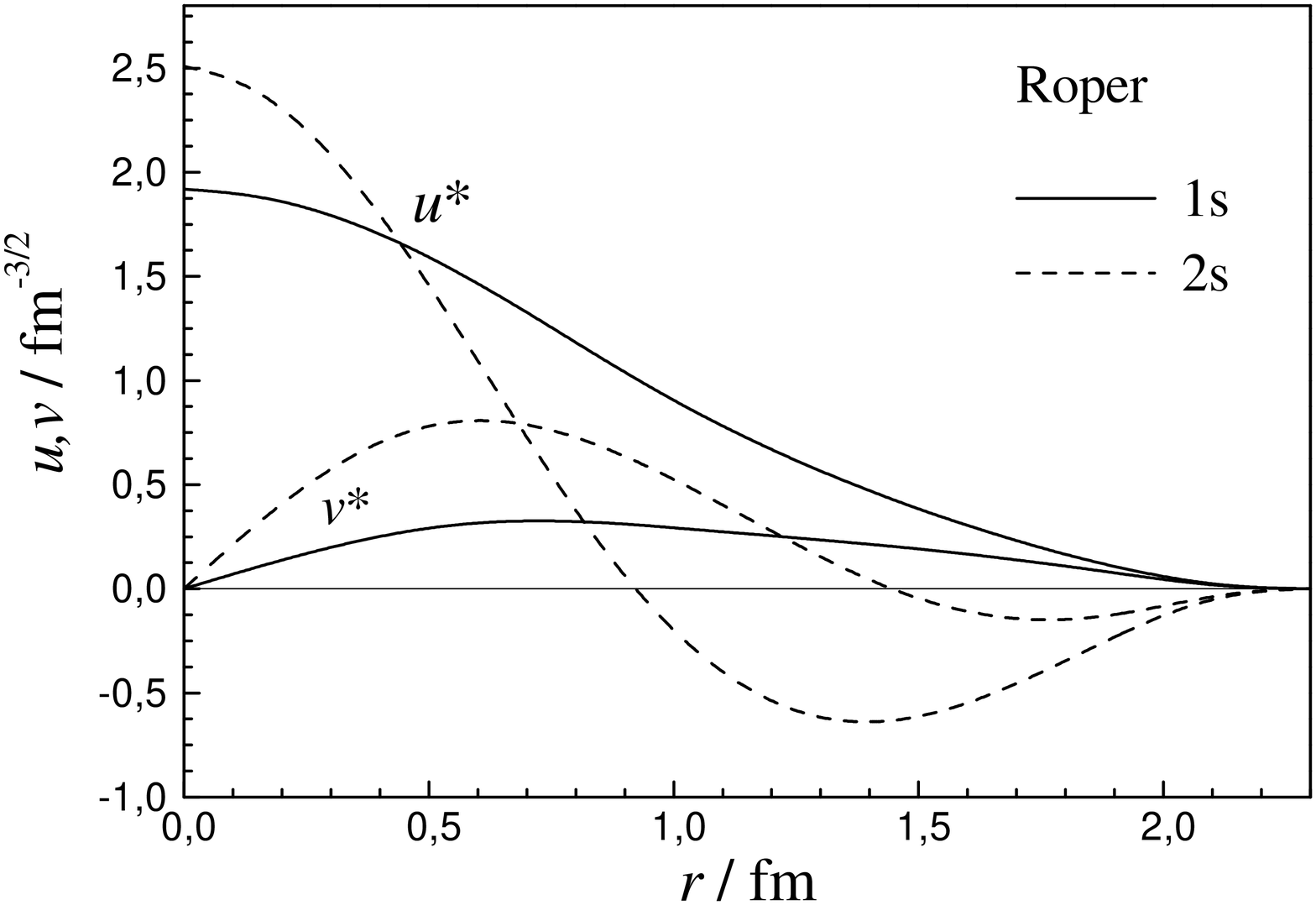,width=6cm}\epsfig{clip=on,
file=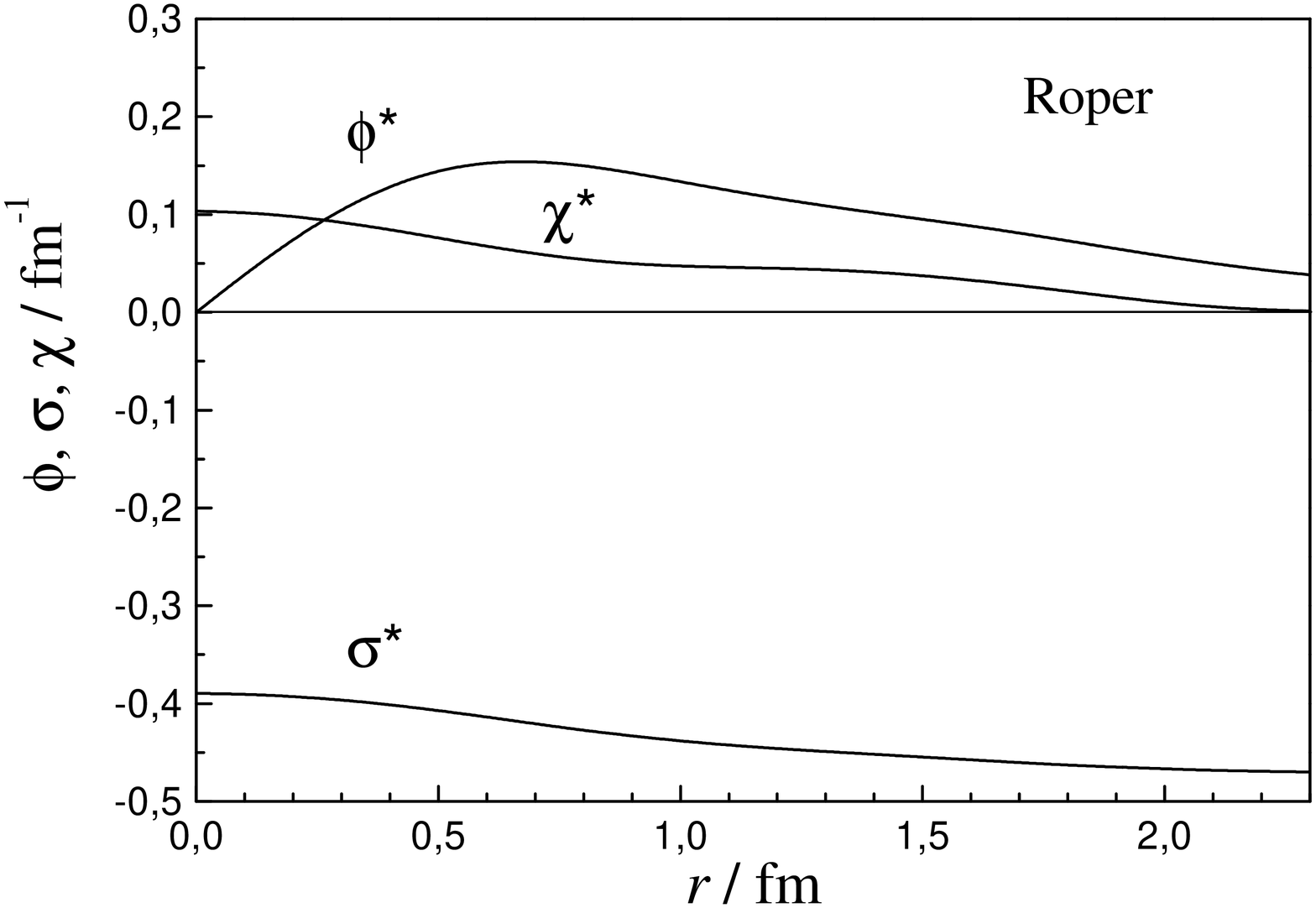,width=6cm}}
\caption{Quark and meson radial wave functions for the  (1s)$^3$
(Nucleon) and (1s)$^2$(2s)$^1$ (Roper) configurations. The vacuum expectation
value of the sigma field is $-f_\pi$. Note that the 
effective quark mass is proportional to the {\em inverse}
of the $\chi$ field. We use the symbol * to denote the Roper radial functions.  
The model parameters are: $M_\chi=1.4$~GeV, $g=0.03$~GeV,
$m_\pi=0.14$~GeV, $f_\pi=0.093$~GeV, $m_\sigma=0.85$~GeV.\vspace*{-0.5cm}} 
\label{fig1j}
\end{figure}

Figure \ref{fig1j} shows the radial profiles for the 
(1s)$^3$ and (1s)$^2$(2s)$^1$ configurations.
Those corresponding to the Roper extend further.  
The strength of the chiral mesons 
is reduced in the Roper in comparison with the nucleon. 
Another interesting feature is the waving shape acquired 
by the Roper chromodielectric field, $\chi^*$. 
A central point in our treatment of the Roper 
is the freedom of the chromodielectric profile, 
as well as of the chiral meson profiles, 
to adapt to a  (1s)$^2$(2s)$^1$ configuration. 
Therefore, quarks in the Roper experience meson fields 
which are different from the meson fields felt by the quarks 
in the nucleon. 
As a consequence,  states (\ref{usa}) are normalized but not mutually
orthogonal. 
They can be orthogonalized taking
\begin{equation}
   |{\rm R}\rangle 
    = {1\over\sqrt{1-c^2}}(|{\rm R}'\rangle - c|{\rm N}\rangle)\;,
\qquad
  c = \langle{\rm N}|{\rm R}'\rangle \;.
\label{orthoR}
\end{equation}
A better procedure results from a diagonalization of the 
Hamiltonian in the  subspace 
spanned by (non-orthogonal) $|{\rm R}'\rangle$ and $|{\rm N}\rangle$:
\begin{equation}
 |\tilde{\rm R}\rangle = c^R_R|{\rm R}'\rangle + c^R_N|{\rm N}\rangle\,, 
\quad
 |\tilde{\rm N}\rangle = c^N_R|{\rm R}'\rangle + c^N_N|{\rm N}\rangle\,. 
\label{gcmss}
\end{equation}

In Table~\ref{tab1j}  the nucleon energies and the nucleon-Roper 
mass splitting are given. 
The absolute value of the nucleon energy is above the experimental
value but it is known~\cite{drago2} that the removal of the 
center-of-mass motion will lower those
values by some 300 MeV (similar correction applies to the Roper). 
On the other hand,
the  nucleon-Roper splitting is small, even in the case of the 
improved state (\ref{gcmss}). 
The smallness of the spitting is probably related with a 
much too soft way of imposing confinement.

\begin{table}[hbt]
\caption{Nucleon energies and nucleon-Roper splittings for two
sigma masses. $E_N$ is the energy of the nucleon state (\ref{usa}), 
$\Delta E$ was obtained using (\ref{orthoR}), $\tilde E$ and 
$\Delta \tilde{E}$ are calculated using the  states (\ref{gcmss}). 
The other model parameters are in the caption of Figure~\ref{fig1j}. 
All values are in MeV. \label{tab1j}}
\begin{center}
\footnotesize
\begin{tabular}{ccccc}
\hline
 & & & &  \\
$m_\sigma$ & $E_N$ & $\Delta E$ & $\tilde{E}_N$ & $\Delta\tilde {E}$ \\
 & & & &  \\  \hline & & & &  \\
\phantom{1}700   &   1249 & 367 & 1235 & 396 \\
1200  &   1269 & 354 & 1256 & 380 \\
 & & & &  \\ 
\hline
\end{tabular}
\end{center}
\end{table}
\vspace*{-0.5cm}

\section{Amplitudes \label{resul}}

\begin{figure}[th]
\centerline{\epsfig{clip=on,file=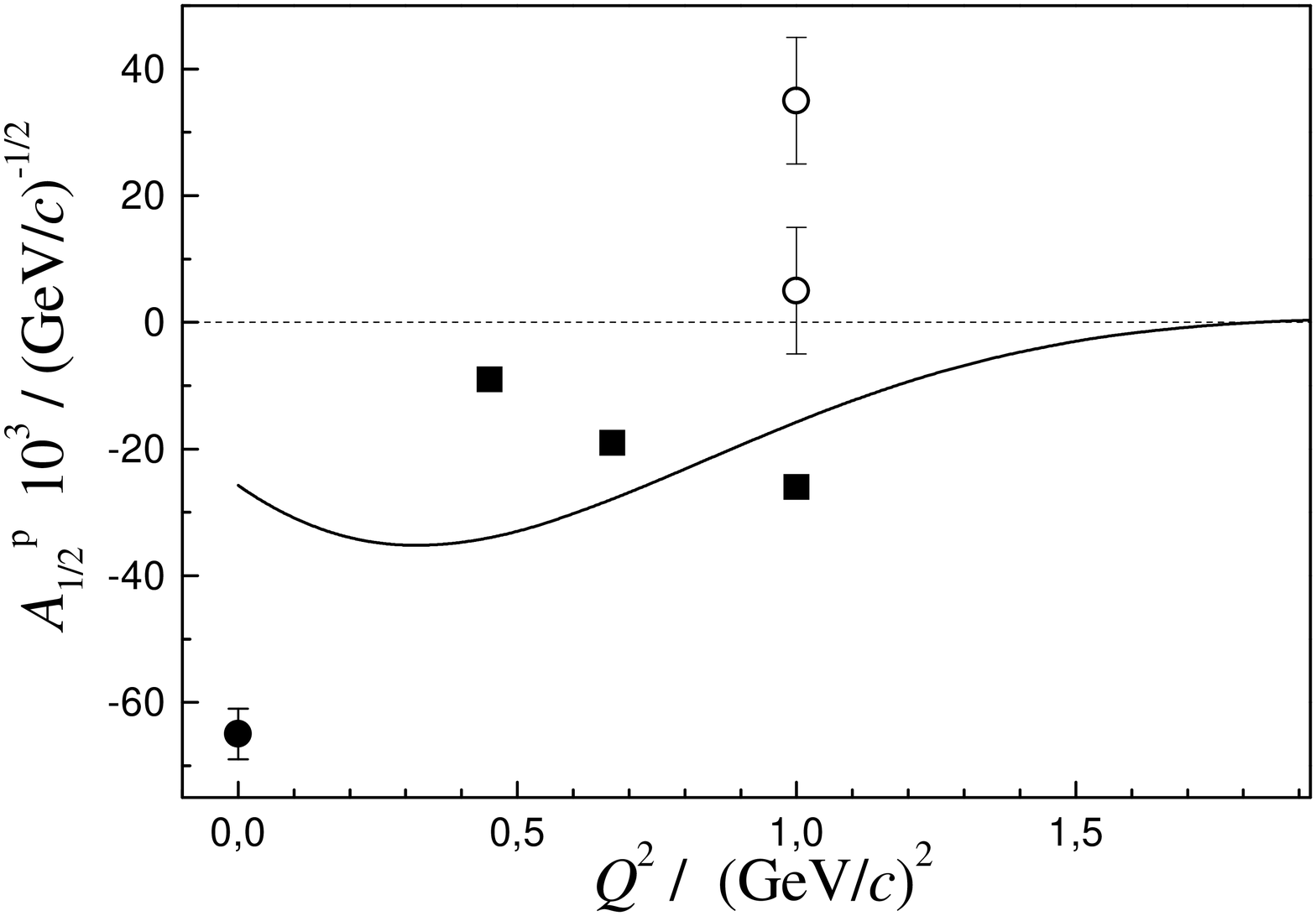,width=6cm}\epsfig{clip=on,
file=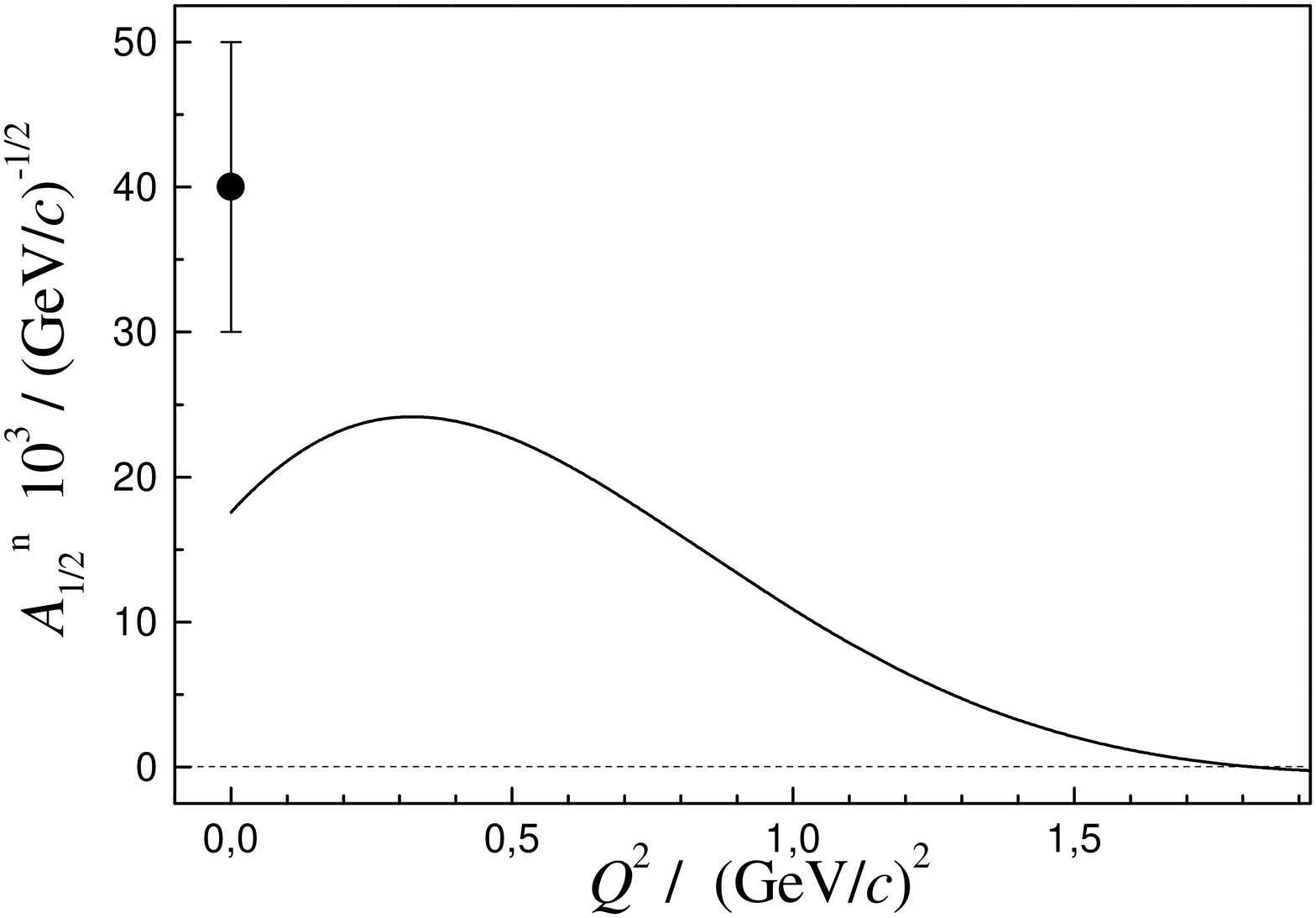,width=6cm}}
\caption[]{Nucleon-Roper transverse amplitudes. 
The experimental points at $Q^2=0$~GeV$^2$ are the estimates of the 
PDG~\cite{PDG}. 
The solid squares~\cite{analy1} and the open circles~\cite{analy2} 
result from the analysis of electroproduction 
data.} \vspace*{-0.5cm}
\label{fig2j}
\end{figure}

Our results for the transverse helicity amplitudes are shown in 
Figure~\ref{fig2j} for the parameter set used for Figure~\ref{fig1j}.
The experimental values at the photon point 
are the PDG most recent estimate~\cite{PDG} 
$A_{1/2}^p=-0.065\pm 0.004$~(GeV/$c$)$^{-1/2}$ and  
$A_{1/2}^n= 0.040\pm 0.010$~(GeV/$c$)$^{-1/2}$. 
The pion contribution to the charged states only accounts for 
a few percent of the total amplitude. 
The discrepancies at the photon point can be
attributed to a too weak pion field,
which we already noticed in the calculation 
of nucleon magnetic moments~\cite{drago2} and
of the electroproduction of the $\Delta$~\cite{delta1}.
Other chiral models~\cite{tiator}
predict a stronger pion contribution which enhances
the value of the amplitudes.
If we calculate perturbatively the leading pion contribution
we also find a strong enhancement at the photon point; however,
when we properly orthogonalize the state with respect to the nucleon,
this contribution almost disappears.

\begin{figure}[hbt]
\centerline{\epsfig{clip=on,file=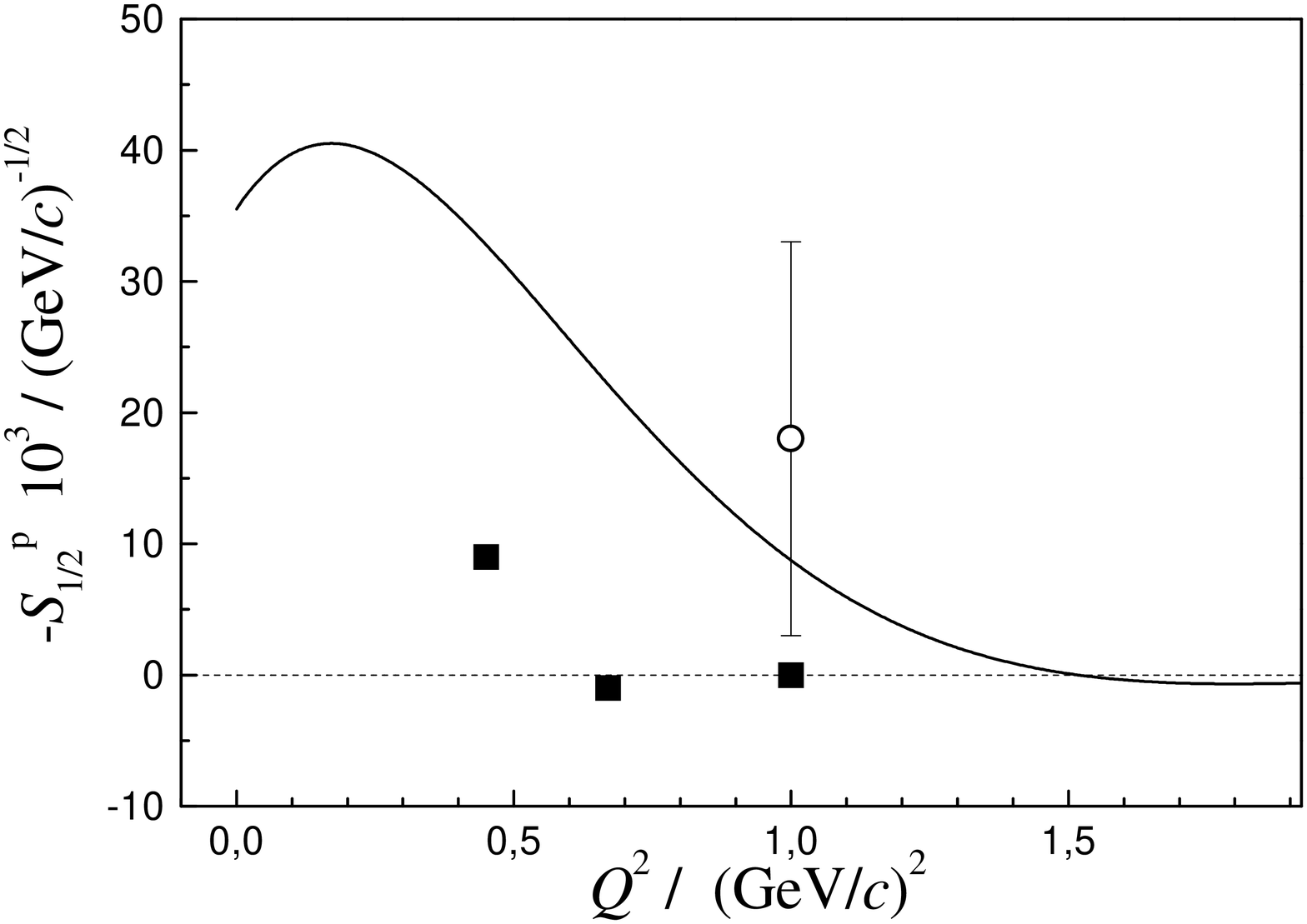,width=6cm}\epsfig{clip=on,
file=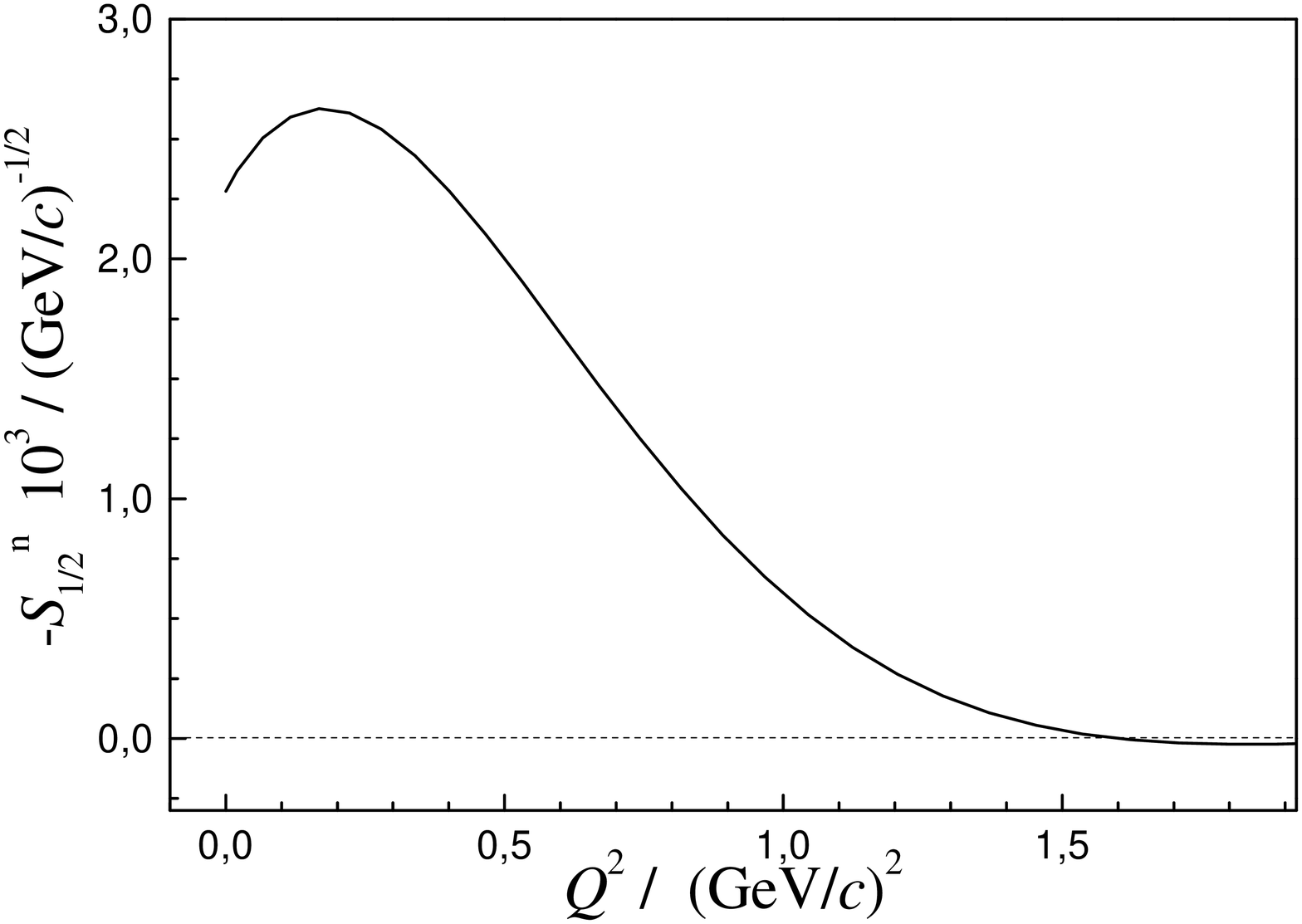,width=6cm}}
\caption{Nucleon--Roper scalar helicity amplitudes (see also 
caption of Figure~\ref{fig2j}).}
\label{fig3j}
\end{figure}

In Figure~\ref{fig3j} we present 
the scalar amplitudes.
For the neutron no data are available which prevents 
any judgment of the quality of our results.

In CQM calculations~\cite{Capstick,Cardarelli,Cano}
which incorporate a consistent relativistic treatment of
quark dynamics,
the amplitudes change the sign around $Q^2\sim 0.2$--$0.5$~(GeV$/c)^2$.
The amplitudes with this opposite sign remain large at relatively high
$Q^2$, though,
as shown in~\cite{Cardarelli,Cano}, the behavior at high $Q^2$
can be substantially reduced 
if either corrections beyond the simple Gaussian-like ansatz
or pionic degrees of freedom are included in the model.
Other models, in particular those including exotic (gluon) states,
do not predict this type of behavior~\cite{Li}.
The present experimental situation is unclear.
Our model, similarly as other chiral models~\cite{tiator,dong},  
predicts the correct sign at the photon point, while
it does not predict the change of the sign at low  $Q^2$.
Let us also note that with the inclusion of a phenomenological
three-quark interaction Cano {\em et al.}~\cite{Cano} shift the change
of the sign to $Q\sim1$~(GeV$/c)^2$ beyond which, in our opinion,
predictions of low energy models become questionable anyway.

\section{Meson excitations\label{mesons}}

The ansatz (\ref{usa}) for the Roper represents the breathing mode 
of the three valence quarks with the fields adapting to the
change of the source.
There is another possible type of excitation in which
the quarks remain in the ground state
while the $\chi$-field and/or the $\sigma$-field  oscillate.
The eigenmodes of such vibrational states are determined by
quantizing small oscillations of the scalar bosons 
around  their expectation values in the ground state~\cite{Bled}.
We have found that the effective potential for such modes
is {\em repulsive\/} for the $\chi$-field and {\em attractive\/} 
for the $\sigma$-field.
This means that there are no glueball excitations in which the 
quarks would act as spectators: the $\chi$- field oscillates
only together with the quark field.
On the other hand, the effective $\sigma$-meson potential supports 
at least one bound state with the energy $\varepsilon_1$ of 
typically 100~MeV below the $\sigma$-meson mass. 

We can now  extend the ansatz  (\ref{orthoR}) by introducing
\begin{equation}
  |{\rm R}^*\rangle = c_1|{\rm R}\rangle 
       + c_2\tilde{a}^\dagger_\sigma|N\rangle\;,
\label{gRstate}
\end{equation} 
where $\tilde{a}^\dagger_\sigma$ is the creation operator 
for this lowest vibrational mode.
The coefficients $c_i$ and the energy are determined 
by solving the (generalized) eigenvalue problems 
in the $2\times2$ subspace.
The lowest energy solution is the Roper
while its orthogonal combination could be attributed
to the $N(1710)$, provided the $\sigma$-meson mass
is sufficiently small.
In such a case the latter state is described as
predominantly the $\sigma$-meson vibrational mode
rather than the second radial excitation of quarks.
This would manifest in very small production amplitudes
since mostly the scalar fields are excited.

The presence of $\sigma$-meson vibrations 
is consistent with the recent phase shift
analysis by Krehl {\em at al.}~\cite{Krehl} who found that
the resonant behavior in the P$_{11}$ channel can be explained
solely through the coupling to the $\sigma$-N channel.
In our view, radial excitations of quarks are needed in order
to explain relatively large electroproduction amplitudes,
which would indicate that the $\sigma$-N channel couples to all 
nucleon $\half^+$ excitations rather than be concentrated
in the Roper resonance alone.

\vskip0.2cm 
This work was supported by FCT (POCTI/FEDER), Portugal, and by 
The Ministry of Science and Education of Slovenia.
MF acknowledges a   grant from GTAE (Lisbon),
which made possible his participation in EMI2001.

\end{document}